\documentclass[twocolumn,showpacs,preprintnumbers,amsmath,amssymb,superscriptaddress,nofootinbib]{revtex4-1}

\usepackage[english]{babel}
\usepackage{amsmath}
\usepackage{graphicx}% Include figure files
\usepackage{dcolumn}% Align table columns on decimal point
\usepackage{bm}% bold math
\usepackage{amscd}
\usepackage{epstopdf}
\usepackage{hyperref}
\usepackage[utf8x]{inputenc}
\usepackage{textcomp}%freccie
\usepackage{natbib}

\usepackage{lipsum}% for dummy text
\begin{document}

\title{The two-particle problem in comb-like structures}

\author{Elena Agliari}
\email[]{agliari@mat.uniroma1.it}
\affiliation{Dipartimento di Matematica, Sapienza Universit\`a  di Roma}

\author{Davide Cassi}
\email[]{davide.cassi@fis.unipr.it}
\affiliation{Dipartimento di Fisica, Universit\`a  di Parma}

\author{Luca Cattivelli}
\email[]{luca.cattivelli@sns.it}
\affiliation{Scuola Normale Superiore, Pisa, Italy}

\author{Fabio Sartori}
\email[]{fabio.sartori@brain.mpg.de}
\affiliation{Max Planck Institute for Brain Research, Frankfurt, Germany}

\date{\today}

\begin{abstract}

Encounters between walkers performing a random motion on an appropriate structure can describe a wide variety of natural phenomena ranging from pharmacokinetics to foraging. On homogeneous structures the asymptotic encounter probability between  two walkers is  (qualitatively) independent of whether both walkers are moving or one is kept fixed. On infinite comb-like structures this is no longer the case and here we deepen the mechanisms underlying the emergence of  a finite probability that two random walkers will never meet, while one single random walker is certain to visit any site.  In particular, we introduce an analytical approach to address this problem and even more general problems such as the case of two walkers with different diffusivity, particles walking on a finite comb and on arbitrary bundled structures, possibly in the presence of loops. Our investigations are both analytical and numerical and highlight that, in general, the outcome of a reaction involving two reactants on a comb-like architecture can be strongly different according to whether both reactants are moving (no matter their relative diffusivities) or only one, and according to the density of short-cuts among the branches. 
\end{abstract}

\maketitle

\section{Introduction}

Random walks (RWs) constitute the basic model for non-deterministic motion and their applications range from biology to economics (e.g., see \cite{Weiss-1994,Klafter-2011,berg1993random,Gillespie-2013,havlin,FP-2014}).
%
%they are applied in a wide range of research areas, such as crystallography \cite{cristal,zumofen1982energy,PhysRevLett.94.080601}, biology \cite{biology,he2008random,jeon2011vivo,barkai2012strange} and even economics to study stock market fluctuations \cite{mantegna}. RWs can model the Brownian motion of a particle immersed in a fluid \cite{fluid}, the thermal motion of electrons in a metal \cite{electrons}, the bacterial motion \cite{bacteria} or the spreading of diseases in dense populations \cite{diseases}, just to cite a few.
%
According to the phenomenon that one aims to model, RWs can be embedded on different structures (e.g., mimicking crystalline solids, glasses, polymers, or social networks \cite{Newman-2010}), whose topology, mathematically described by graphs \cite{burioni1}, strongly affects the diffusive behavior. 
\begin{figure}[tbh]
\noindent \begin{centering}
\includegraphics[width=0.45\textwidth]{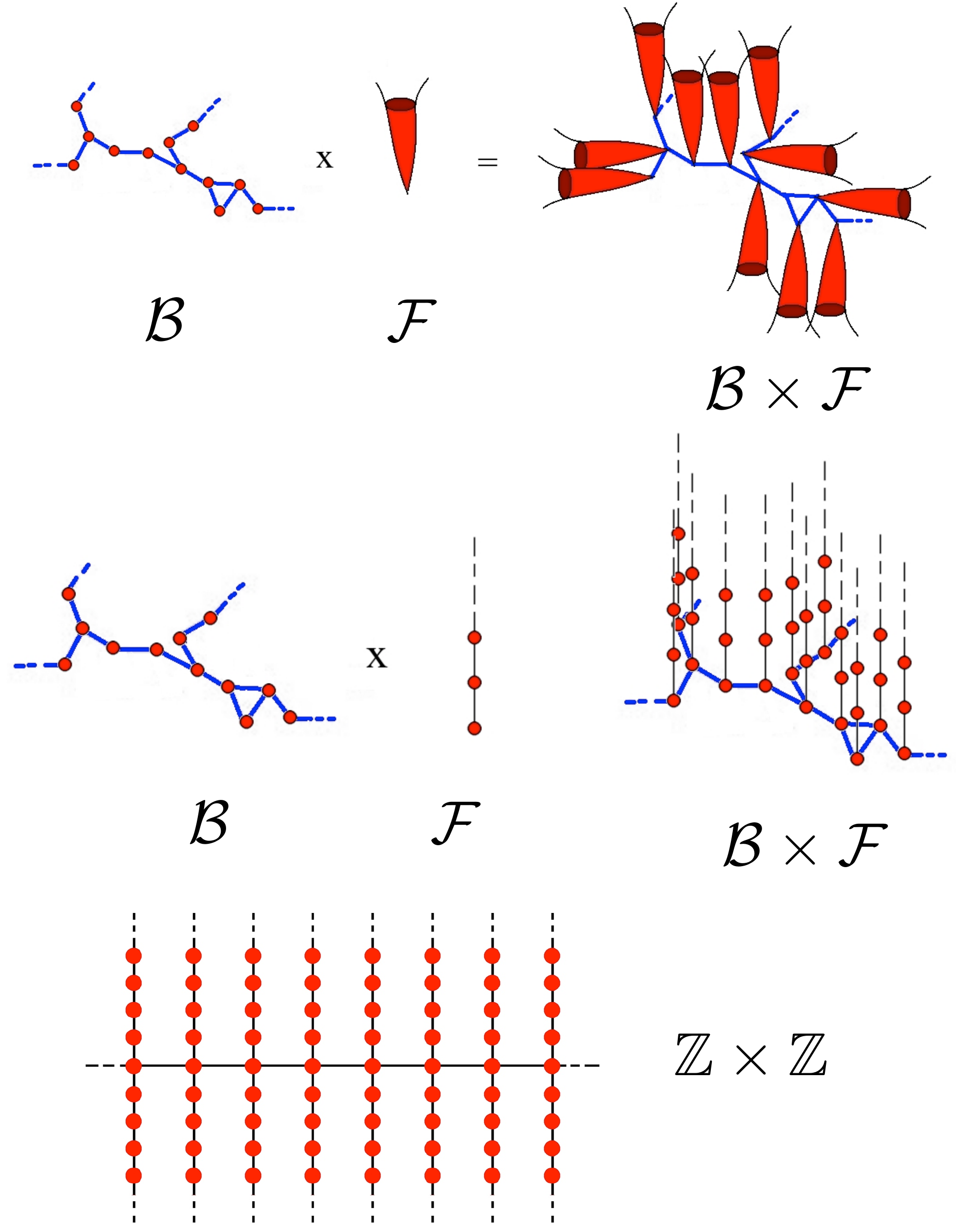}
\par\end{centering}
\caption{ (Color online) Upper panel: generic branched structure obtained by attaching to each site of the arbitrary base $\mathcal{B}$ an arbitrary fiber $\mathcal{F}$. Middle panel: generic comb obtained by attaching to each site of the arbitrary base $\mathcal{B}$ a linear chain. Lower panel: simple $2$-dimensional comb obtained by taking as base a linear chain and attaching a linear chain to each site of the base.}
\label{fig:comb}
\end{figure}
In particular, natural structures, such as macromolecules and disordered materials, often exhibit a tree-like architecture (see e.g., \cite{7,8,9,10}). A very versatile and interesting model for such systems is given by \emph{combs} (see Fig.~\ref{fig:comb}), which
can be defined as discrete structures obtained by joining to each point of a ``base'' graph a linear chain, or, more generally an arbitrary ``fiber'' graph (the latter case is often referred to as ``branched structure'').
The random walk problem in simple combs, where both fibers and base are linear chains, has been extensively investigated in the last years (see e.g., \cite{ASCC-PRE2015, agliari_blumen, Cassi_Campari_2012, Zhang-PRE2010, havlin, mendez2015mesoscopic, iomin2015destruction, ribeiro2014investigating, benichou2015diffusion}).

In this work we focus on the encounter of two RWs on comb-like structures. Collisions between two random walkers can be seen as the basic process underlying diffusion-limited (or diffusion-controlled) reactions (see, e.g., \cite{havlin,redner,FP-2014,reaction3}), where reactions occur upon reactants encounter and the time-scale for reaction is much shorter than the characteristic time for the two walkers to meet. 
%This can describe a wide variety of natural phenomena, including prey-predator interaction \cite{prey1,prey2}, chemical reaction kinetics \cite{chemical,zumofen1985concentration,velonia2005single}, foraging \cite{foraging1,foraging2} and pharmacokinetics \cite{pharmaco1,pharmaco2}.
This problem is by far non trivial given the topological inhomogeneity of comb-like architectures. In fact, while on homogeneous structures, such as Euclidean lattices, the asymptotic encounter probability between  two walkers is  (qualitatively) independent of whether both walkers are moving or one is kept fixed, in some inhomogeneous structures, such as combs \cite{Krishnapur-ECP2004,Chen-JProb2011}, if both agents move there is a finite probability that they will never meet, while if one stays put and the other moves they eventually meet with certainty \cite{Cassi_Campari_2012}. This property is called \emph{two-particle transience} and it may yield to effects of practical importance; for instance, chemical reactions are favored when either of the reagents is immobilized.

In this paper we outline an effective framework for the analytical investigation of the two-particle problem on comb-like structures. Within such framework we aim to deepen the mechanisms underlying the emergence of the two-particle transience and to address more general problems such as the case of two particles with different diffusivities, particles moving on a finite comb and on arbitrary bundled structures. The problem is further investigated via numerical simulations to corroborate analytical findings and to highlight the robustness of the two-particle transience as new links in the underlying structure are progressively inserted.

In particular, we find that the two-particle transience is preserved as long as both particles are moving (no matter their relative diffusivities), and as long as the walkers spend sufficiently long time on fibers (if the base is recurrent, and a fortiori if the base is transient). 
The two-particle transience is preserved also when an extensive number (yet sublinear in the volume) of bridges is inserted among the branches of a simple comb.

The paper is organized as follows: in Sec.~\ref{sec:comb} we provide an alternative proof of the two-particle transience on combs and, whithin this framework, in Sec.~\ref{sec:further} we address several extensions and generalizations of the problem: in Sec.~\ref{sec:motion} we estimate the probability for the encounter to occur either on the backbone or on a tooth, in Sec.~\ref{sec:velocities} we discuss the case of walkers with different diffusivity, in Sec.~\ref{sec:finite_comb} we deals with finite-size effects, and in Sec.~\ref{sec:bundled} 
we present related results in bundled structures. Then, in Sec.~\ref{sec:ponti} we numerically check the robustness of the result by topologically perturbing the simple comb. 
The phenomenon is further analyzed in Sec.~\ref{sec:The Mapping} from a different perspective, while Sec.~\ref{sec:concl} is left for conclusions and discussions. Finally, Appendices A and B contain some technical tools concerning the relations between the one- and the two-particle problem.

\section{Two random walkers on combs\label{sec:comb}}

In this section we consider the two-particle problem on two-dimensional combs looking for intuitive arguments to show the emergence of the two-particle transience. While this problem has already been treated rigorously \cite{Krishnapur-ECP2004,Chen-JProb2011,Cassi_Campari_2012}, our perspective aims to focus on the key mechanisms underlying the two-particle transience and to possibly extend the phenomenology to more general structures and situations.

\begin{figure}[bt]
\noindent \begin{centering}
\includegraphics[width=0.45\textwidth]{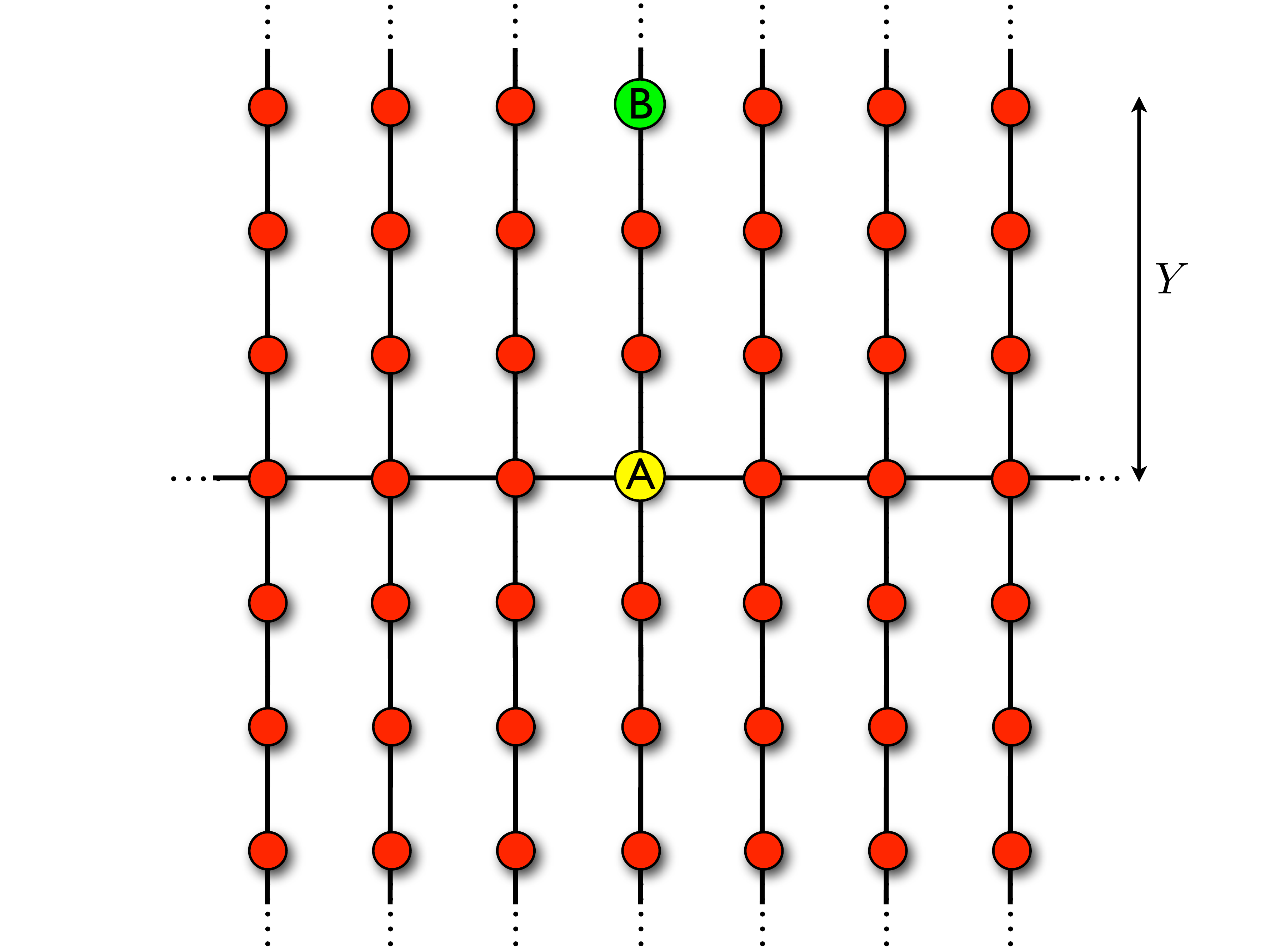}
\par\end{centering}
\caption{ (Color online) Two particles (named A and B, respectively) on the comb sharing the same tooth. In this picture when the walker A reaches the tooth already occupied by the walker B, the latter is at position $Y$ along the side chain, namely, at that time $\Delta_x =0$ and $\Delta_y=Y$.}
\label{two particles}
\end{figure}

\begin{figure}[h]
\noindent \begin{centering}
\includegraphics[width=0.5\textwidth]{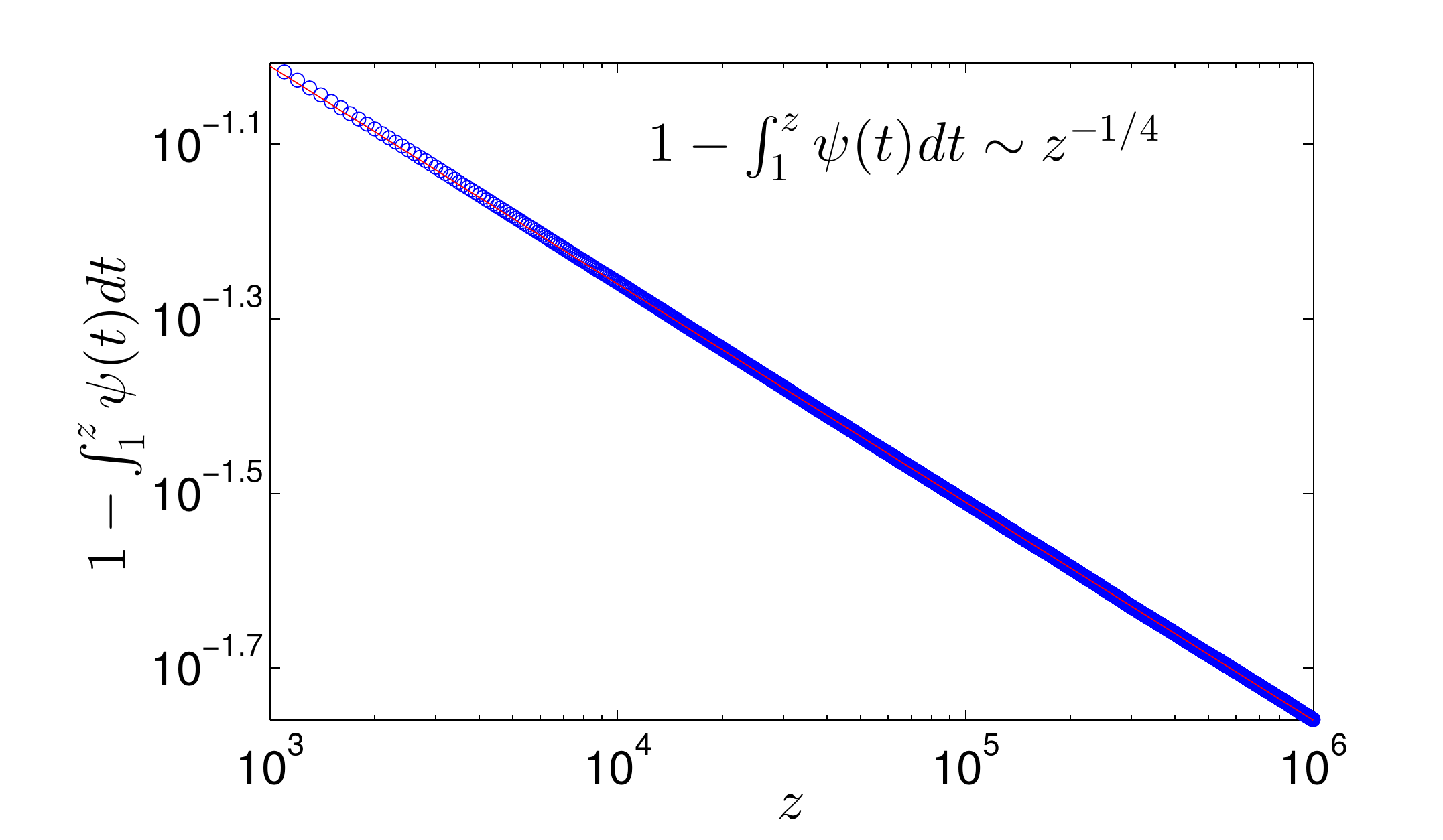}
\par\end{centering}
\caption{ (Color online) In order to check the scaling in Eq.~\ref{eq:Fcomb_a} it is convenient to look at the complementary of the cumulative distribution, namely at the quantity $1-\int_1^z \psi( t) d t$, which represents the probability that two random walkers have not yet shared the same tooth after a time $z$. This quantity is obtained via numerical simulations (bullets) and successfully compared with the power-law $\sim z^{-1/4}$ (solid line) resulting from the estimate in Eq.~\ref{eq:Fcomb_a}.
Numerical simulations are performed on an ``infinite'' comb, where the walkers are initially set on the backbone with relative distance of two sites ($\Delta_x=2$).
At each time step ($t=1,2,3,...$) the two walkers change synchronously their position toward a nearest site selected with equal probability. The underlying ``infinite'' comb, is mimicked by not imposing any boundary conditions and by using a data type for the instantaneous positions whose maximum cannot be reached in the considered time interval.  The latter is fixed by a cut-off in time corresponding to $10^6$. Thus, a simulation stops upon the walkers find themselves on the same tooth at a certain time $t < 10^6$ or whenever the time cut-off is reached.
The results shown here have been averaged over $10^7$ replicas.}
\label{stessa_x}
\end{figure}

We recall that a two-dimensional comb can be obtained by joining to each point of a linear chain (playing as the ``base'') two linear chains (playing as the ``fiber'') as shown in Fig.~\ref{fig:comb} (lower panel). 
%Otherwise stated, two-dimensional comb is obtained attaching at each point of $\mathbb{Z}$ another copy of $\mathbb{Z}$ by its origin.
%
Now, let us consider two walkers, starting from an arbitrary initial position at time $t=0$. % , after a time $t$ occur to be in different teeth without having ever met.
Of course, a necessary condition for meeting is being in the same tooth. Therefore, a useful quantity to look at is the time when both walkers first occur to be on the same tooth \footnote{To fix ideas, in this calculation we are assuming that walkers are starting from nodes belonging to different teeth. Recovering the case where walkers start  from the same tooth just implies sub-leading corrections, as the probability that the walkers eventually share the same tooth is always unitary whatever the initial configuration.}, namely when a walker (say the one named A in Fig.~\ref{two particles}) first enters in the tooth already occupied by the other walker (named B in Fig.~\ref{two particles}). In order to estimate this time we can exploit the translation invariance along the backbone and just focus on the distance $\Delta_x$ between the projections on the backbone of the positions of the two walkers. 
In fact, we expect that the probability $\psi( t )$ for two walkers to be in the same tooth (that is, $\Delta_x=0$) for the first time at time $t$  and the probability $\psi_0( t )$ that a single walker first returns to the original tooth at time $t$ display the same scaling. One can see that the latter scales as $\psi_0( t ) \sim  t^{-5/4}$ \cite{bundled,burioni1} in such a way that 
\begin{equation}\label{eq:Fcomb_a}
\psi(t)\sim t ^{-5/4}.
\end{equation}
This result is deepened in the Appendix \ref{mapping_appendix} and successfully checked via numerical simulations as shown in Fig.~\ref{stessa_x}.

Given that the two walkers share the same tooth, we are interested in their mutual distance $\Delta_y$ along the common tooth. Referring to Fig.~\ref{two particles}, we aim to get the distribution $h(Y)$ where $Y$ is the distance between A and B at the time when A enters the tooth already occupied by B, namely $Y$ is equivalent to the distance of B from the backbone. % minus $1$: $Y=|Y_b|-1$, and so $Y\geq0$.
 The coordinate $Y$ can be treated as a normal random variable with variance scaling linearly with time. % (***Stiamo quindi assumendo che B non si trovi sul backbone, ma in fondo la probabilità che si trovi sul bb è infinitesima***)
In fact, at the arbitrary time $t$, the position of B along a generic tooth \footnote{We recall that we are exploiting the translation invariance along the backbone. Also, the presence of the backbone only introduces a probability $1/2$ that at $Y=0$ the walker does not change its coordinate, namely there is a unitary waiting time \cite{bertacchi,havlin}.} will be distributed as $h(Y,t) \approx  e^{- (Y-Y_0)^{2} /(2 t)} / \sqrt{2\pi t}$, where $Y_0$ accounts for the initial position of B and it can be set equal to zero without loss of generality if we are interested in asymptotic times.
In order to obtain the probability distribution $h(Y)$ of the position $Y$ of the walker B, when A and B first share the same tooth, we need to integrate over $\psi(t)$, namely
\begin{equation}
\begin{split}
h(Y) & =\int \psi( t ) h(Y, t )\,d t\sim\\
     & \int t^{-5/4}\frac{1}{t^{1/2}}\, \exp\bigg(-\frac{Y^{2}}{2 t}\bigg)\,d t \sim\frac{1}{Y^{3/2}}.\label{eq: salti1}
\end{split}
\end{equation}
Therefore, every time the two walkers begin to share the same tooth, their relative distance $\Delta_y$ along the tooth is a random variable following the probability distribution $h(Y) \sim Y^{-3/2}$. Otherwise stated, the relative distance along $y$ follows a long-tail distribution $h(Y)\sim Y^{-\mu-1}$ with $\mu=1/2$, in such a way that, in the average, they are at an infinite distance. The last remark already provides an intuitive argument for understanding the origin of the two-particle transience.

Now, referring again to Fig.~\ref{two particles}, as the walkers A and B occur to be on the same tooth, the former can either move on a different tooth without having the chance to encounter the latter, or they can encounter before the walker A escapes from the common tooth.

 This problem can be recast into a single walker moving in a
semi-infinite chain in the presence of a target at a distance $Y$, and we are interested in the probability for the walker to visit the target before returning to the origin of the chain. This case was addressed in \cite{Majumdar} where the authors found that this probability is given by $2-\frac{4}{\pi} \arctan(Y) $.
Exploiting this result we can pose that, in the limit $Y \to \infty$, the probability $a(Y)$ for the walker A to encounter B before returning to the backbone scales as
\begin{equation} \label{eq:asterisco}
a(Y) \sim \frac{1}{Y}.
\end{equation}

% This problem can be recast into a single walker moving in a semi-infinite chain in the presence of a target at a distance $Y$, and we are interested in the probability for the walker to visit the target before reaching the origin. This probability, referred to as $a(Y)$, scales as $a(Y)=2-\frac{4}{\pi} \arctan(Y/Y') $ \cite{Majumdar}, where $Y^{\prime}$ is the initial position of the walker A on the tooth. In the limit $Y^{\prime}/ Y \to 0$, we find that 
 
%\begin{equation} \label{eq:asterisco}
%a(Y) \sim \frac{1}{Y}.
%\end{equation}  
Therefore, the encounter probability for the two walkers on the comb is ultimately controlled by two quantities:
\begin{itemize}
\item{the distribution $h(Y)$ for the relative distance $Y$ as the walkers are on the same tooth;}
\item{the encounter probability  $a(Y)$ when the walkers are in the same tooth.}
\end{itemize}

The whole process can be seen as a L\'{e}vy flight \cite{PhysRevA.35.3081,zumofen1989levy,klafter1990levy,metzler2004restaurant} on a linear chain in the presence of absorbing traps distributed according to $a(Y)$. Each jump of the L\'evy flight corresponds to the two walkers sharing the same tooth\footnote{More precisely, we should consider also the waiting time $\psi(t)$ between two consecutive jumps. Indeed, the jump sizes and the waiting times are coupled and the process is better described in terms of a Coupled Continuous Time Random Walk (CCTRW). However, as shown in \cite{CASC-PRE2015}, the absorption probability depends only on $a(Y)$ and $h(Y)$ and not on the distribution $\psi(t)$. Therefore, the description of the process in terms of L\'evy flights is perfectly workable.}. At each jump the L\'evy flight can either be absorbed or move on.
 If the L\'{e}vy flight is eventually absorbed with probability $1$, then the particles surely meet. This model was studied in detail in \cite{CASC-PRE2015}, where it is shown that the L\'{e}vy flight characterized by a jump distribution  $h(\xi) \sim \xi^{-\mu -1}$ in the presence of traps with distribution scaling as $a(\xi) \sim \xi^{-\alpha}$ has a finite probability of never being absorbed when the displacement exponent $\mu$ (in our case $\mu=1/2$) is lower than the absorption exponent $\alpha$ (in our case $\alpha=1$).
Since here this condition is fulfilled, we recover the two-particle transience on combs, that is, the two particles have a finite probability of never meeting, regardless of their starting position.

\section{Further characterizations and extensions}\label{sec:further}
In this section we exploit the framework introduced in the previous section in order to get information on the spatial distribution of encounters and to extend the emergence of the two-particle transience on more general structures and situations.

\subsection{Spatial distribution of the encounters}\label{sec:motion}

In this subsection we investigate the spatial distribution for the location of the encounters, trying to estimate the probabilities $P_{\textrm{backbone}}(t)$ and $P_{\textrm{tooth}}(t)$ that an encounter (if any) between the two walkers occurs in the backbone or in a tooth, respectively. 

The position of a walker can be univocally determined by specifying its projection on the backbone and its height along the tooth. Let us denote with $(X_{1}, Y_{1})$ and with $(X_{2}, Y_{2})$ such coordinates for particle $1$ and for particle $2$, respectively.  
The evolution of the coordinates along $x$ can be seen as a continuous-time random walk on a one-dimensional lattice, while the evolution of the coordinates along $y$ can be seen as normal diffusion \cite{havlin,bertacchi}. 
%Indeed, this motion can be seen as a simple random walk with finite waiting times when $y=0$, which are the moments when the walker moves along the backbone. These pauses during the motion does not have affect the kind of motion \cite{burioni1}.  
As a consequence, the coordinates $Y_{1}$ and $Y_{2}$ can be treated as normal random variables with variance scaling linearly with time. %\footnote{The presence of the backbone only introduces a probability $1/2$ that at $Y=0$ the walker does not change its coordinate, namely there is a unitary waiting time \cite{bertacchi}.}
The relative distance $\Delta_{y}=Y_{1}-Y_{2}$ and the position of the center of mass $Y_{\textrm{cm}}=(Y_{1}+Y_{2})/2$ along $y$ are as well normal random variables, being a difference and a sum, respectively, of Gaussian variables.
More precisely, we can state that $Y_{\textrm{cm}}$ has a Gaussian evolution:
\begin{equation}\label{eq:Ycm}
P(Y_{\textrm{cm}},t)= \mathcal{N}(Y_{\textrm{cm}}(0),D t),
\end{equation}
where $D \in \mathbb{R}^{+}$ is the diffusivity constant \footnote{Notice that here we are implicitly assuming that the two walkers start in a configuration with $Y_{\textrm{cm}} =0$. This does not affect the asymptotic behaviour of $P(Y_{\textrm{cm}}=0,t)$ given in Eq.~\ref{eq:4}.}.
% (e.g., see \cite{BC-JPA2005}).}.
\newline
Now, the encounter in the backbone is characterized by $Y_{\textrm{cm}}=0$ and $\Delta_{y}=\Delta_{x}=0$. According to Eq.~\ref{eq:Ycm} and to the arguments discussed in Appendix \ref{app}\footnote{In Appendix \ref{app} we show that the probability $P(\Delta_x=\Delta_y=0,t)$ for the encounter between two walkers and the probability of return to the origin for a single walker display the same scaling with time. For the latter is scaling is known to be $\sim t^{-3/4}$ \cite{burioni1}.}, these conditions correspond to
\begin{eqnarray}
\label{eq:4}
P(Y_{\textrm{cm}}=0,t) &\sim& t^{-1/2},\\
\label{eq:5}
P(\Delta_x=\Delta_y=0,t) &\sim& t^{-3/4},
\end{eqnarray}
and they have to be satisfied at the same time. Since the potions along $x$ and $y$ are asymptotically uncorrelated \cite{havlin} and the walkers move independently, we can write
\begin{eqnarray}
\nonumber
P_{\textrm{backbone}}(t) 	& = 	& P(\Delta_x=\Delta_y=0,t)\,P(Y_{\textrm{cm}}=0,t)\\ 
\label{eq:backbone}
&\sim	& t^{-3/4}\,t^{-1/2}\sim t^{-5/4}.
\end{eqnarray}

\begin{figure}[t]
\noindent \begin{centering}
\includegraphics[width=0.4\textwidth]{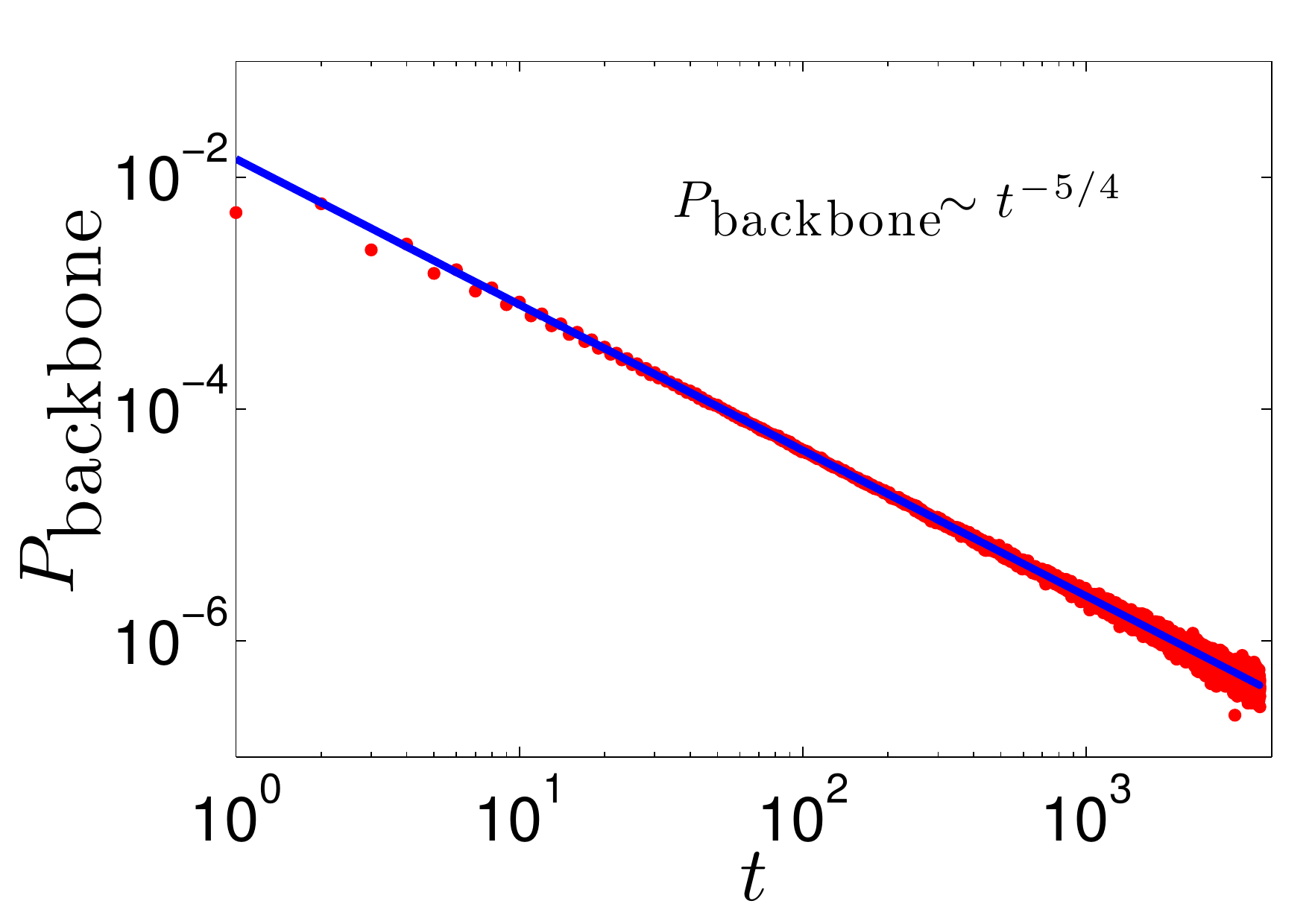}
\includegraphics[width=0.4\textwidth]{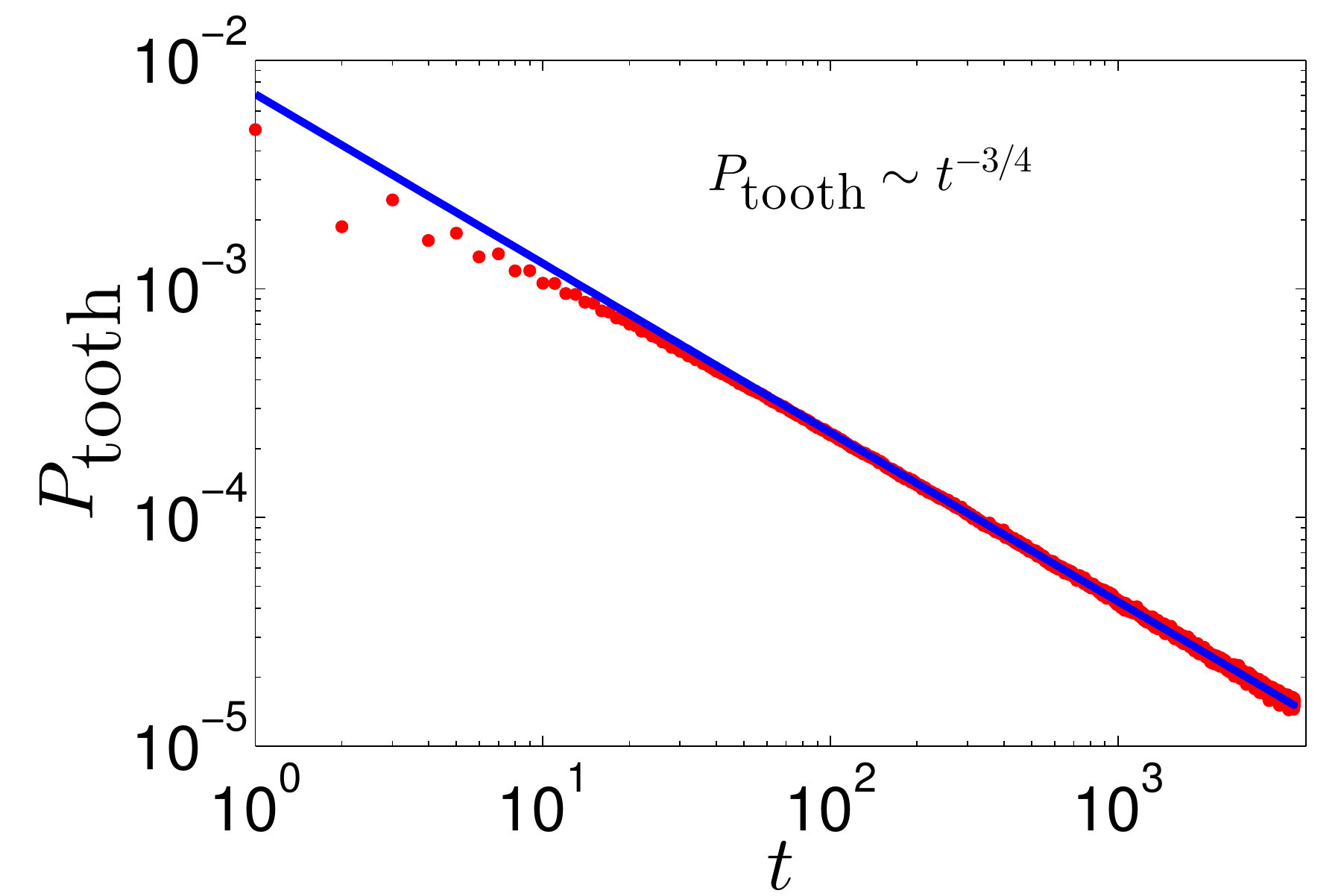}
\par\end{centering}
\caption{ (Color online) Probability $P_{\textrm{backbone}}(t)$ that the two walkers encounter in a site belonging to the backbone (upper panel)
and probability $P_{\textrm{tooth}}(t)$ that the two walkers encounter in a site belonging to teeth (lower panel). Results from numerical simulations (bullets) are successfully compared with analytical estimates (solid line) according to Eqs.~\ref{eq:backbone} and \ref{eq:tooth}, respectively.  
In the numerical simulations the walkers are initially set on the backbone with relative distance $\Delta_x=2$ and at each time step ($t=1,2,3,...$) they change synchronously their position toward a nearest site selected with equal probability. A simulation stops as a time threshold $4 \times 10^3$ is reached and the size of the comb is taken large enough that, for this temporal cut-off, the walkers do not realize its finiteness. We repeat the simulation $10^7$ times and for each realization we keep track of the time step $\tau$ when walkers possibly occur to encounter on the backbone (upper panel) or on a tooth (lower panel). The final distributions are then obtained as histograms over $\tau$. Note that in a single realization, there may be more than one encounters and therefore a single realization may return several values for $\tau$. }
\label{backbone}
\end{figure}

Let us now consider the probability that the encounter occurs on a tooth. This can be evaluated as the asymptotic behaviour of the encounter probability minus the asymptotic behaviour of the probability of encountering in the backbone:
\begin{eqnarray}
\nonumber
P_{\textrm{tooth}}(t)		& =		&P(\Delta_x = \Delta_{y}=0, t)\left[1-P(Y_{\textrm{cm}}=0,t)\right]\\
\label{eq:tooth}
&\sim& t^{-3/4}-t^{-5/4}\sim t^{-3/4}.
\end{eqnarray}
The results in Eq.~\ref{eq:backbone} and in Eq.~\ref{eq:tooth} are successfully checked numerically, as shown in Fig.~\ref{backbone}.

Remarkably, we have evidenced that encounters in the backbone are asymptotically negligible with respect to encounters in teeth. Moreover, by integrating $P_{\textrm{backbone}}(t)$ and $P_{\textrm{tooth}}(t)$ over time, we get an estimate for the average number of encounters occurring on the backbone and on the teeth, respectively; using the expressions in Eqs.~\ref{eq:backbone} and \ref{eq:tooth}, one can see that in the former case the average number of encounters is finite, while in the latter it is infinite.

\subsection{Walkers with different diffusivities } \label{sec:velocities}

In this subsection we analyze the case of walkers with different diffusivities.
Being time and space discrete, this equals to say that  particles A and B take a step only at times multiple of two natural numbers $n_A$ and $n_B$. 
Of course, when one of the two particles is static (say, $n_A=\infty$ and $n_B < \infty$) we recover the one-particle problem and the encounter occurs with probability one \cite{d-comb}, while when $n_A=n_B$  we recover the standard two-particle problem with the emergence of the two-particle transience.  One may therefore wonder whether the transition between the two-particle recurrence  and the two-particle transience  occurs at any finite value of the ratio $n_A/n_B$.

Referring to Fig.~\ref{two particles}, the first quantity to look at is the distribution $\psi(\tilde{t})$, where $\tilde{t}$ is the time when both walkers first occur to be on the same tooth. Since the number of steps for unit time is now rescaled by a finite constant $n (n_A, n_B)$ which depends on $n_A$ and $n_B$, the distribution $\psi(\tilde{t})$ is simply rescaled by the same factor leaving the asymptotic behaviour unaffected:
\begin{equation}\label{eq:Fcomb}
\psi(\tilde{t})\sim (n\, \tilde{t}) ^{-5/4}\sim  \tilde{t} ^{-5/4}.
\end{equation}
In the time interval $\tilde{t}$, the position $Y$ of B along the tooth  is distributed as $h(Y,\tilde{t}) \approx  e^{- (Y-Y_0)^{2} /(4 D_B \tilde{t})} / \sqrt{4\pi D_B \tilde{t}}$, where $D_B$ is the diffusivity of $B$ along the teeth. 
In order to obtain the probability distribution $h(Y)$, when A and B share the same tooth, we need to integrate over $\psi(\tilde{t})$, namely
\begin{equation}
\begin{split}
h(Y) & =\int \psi( \tilde{t} ) h(Y, \tilde{t} )\,d \tilde{t}\sim\\
     & \int \tilde{t}^{-5/4}\frac{1}{\sqrt{4 \tilde{t} D_B}}\, \exp\bigg(-\frac{Y^{2}}{4 \tilde{t} D_B}\bigg)\,d \tilde{t} \sim\frac{1}{Y^{3/2}},\label{eq: salti h(x)}
\end{split}
\end{equation}
in analogy with Eq.~\ref{eq: salti1}.

Moreover the encounter probability for the walker A to encounter the target B before leaving the tooth scales as \cite{Majumdar}:
\begin{equation}
a(Y)\sim \sqrt{\frac{D_B}{D_A}} \frac{1}{\arctan\left(\frac{D_B}{D_A}\right)}\frac{1}{Y}\sim \frac{1}{Y},
\end{equation}
in analogy with Eq.~\ref{eq:asterisco}, independently of the values of the two diffusivities (provided that they are both finite and non null).  

We conclude that the transition between two-particle transience and two-particle recurrence (that is, when the walkers surely meet) is ``trivial" as it occurs at $n=0$ (or $n = \infty$). This result is deepened in the Appendix \ref{mapping_appendix}.

\subsection{Two random walkers on a finite comb}\label{sec:finite_comb}

In this section we traslate the framework discussed in Sec.~\ref{sec:comb} to the case of finite sized combs. In fact, the (possible) encounter process can again be split in two phases: ``approaching'' (which ends upon the two walkers share the same tooth) and ``tackling'' (which ends upon the two walkers either meet on the common tooth or cease to share the same tooth). However, dealing with finite sizes, it will be more convenient to focus on the average of the observables rather than on their distribution. 

%NOTA: Questa sezione andrebbe resa piu' omogenea (in notazione e approccio) con la parte precedente\\
%Secondo me, il legame tra questa sottosezione e il nostro framework è il seguente:\\
%Usiamo sempre lo stesso approcio: guardiamo il tempo affinche le due particelle si trovino nello stesso dente, poi guardiamo la probabilità di incontro quando sono nello stesso dente, e se non si incontrano, si ripetono questi due passaggi fino al momento del meeting.

Before proceeding it is worth stressing that, by definition, the two-particle transience is a property emerging in the thermodynamic limit, yet real structures are necessary finite and it is therefore important to see whether finite-dimensional structures also keep any track of such a property. In particular, the case of finite combs was studied in \cite{agliari_blumen}, and it was shown that, in finite combs, the encounter between two particles is ``slow'', namely the characteristic time for two random walkers to first meet is qualitatively larger than the characteristic time for one single particle to first reach a fixed target. More precisely, if one walker stays fixed in a given site of the backbone of a $2$-dimensional comb and the other one moves throughout the comb starting from the same site, the mean encounter time scales with $L^{2}$. On the other hand, when both walkers are moving, starting from the same position in the backbone, the average encounter time %scales as $f(L)\sim O(L^2)$,
increases more than quadratically, that is $\frac{f(L)}{L^2}\to \infty$ (but $\frac{f(L)}{L^3}\to 0$). The latter result was found numerically \cite{agliari_blumen}, while here we want to recover this time dilation analytically, exploiting the framework introduced in Sec.~\ref{sec:comb}.

%More precisely, as for the case of an infinite comb discussed in Sec.~\ref{sec:comb}, we study the encounter probability between the two walkers dividing the motion along the backbone from the motion along the teeth.  
First, we consider the average time for the walkers to be in the same tooth when they start from the same position on the backbone. The motion along the backbone is  a continuous-time random walk on a finite chain with mean waiting time $\sim L$ \cite{havlin}, which is the mean time spent wondering along a tooth. 
The mean time for the walkers to occupy the same position on a finite chain is $\sim L$ \cite{redner}, then, considering the effect of the waiting times, the average time for sharing the same tooth is $\tau_1\sim L^2$. Now, even if walkers occur to be on the same tooth, there is no certainty for the meeting, because a walker can leave the tooth before encountering the other walker. If they do not encounter and one of the two walkers leaves the tooth, it will need another mean time $\tau_1$ to share the same tooth, and so on. 

\begin{figure}[tb]
\includegraphics[width=0.4\textwidth]{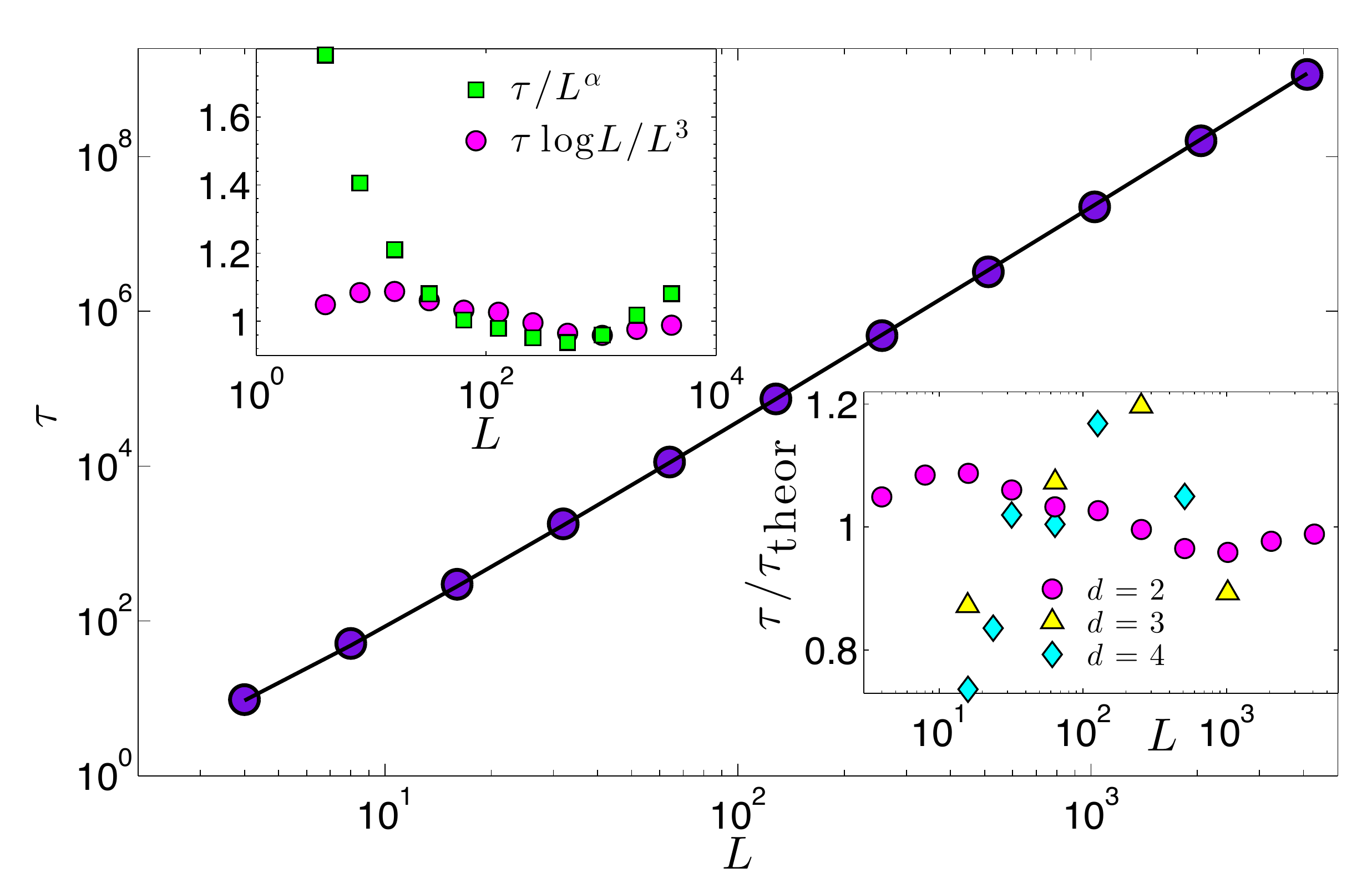}
\caption{ (Color online) Main plot: Mean encounter time $\tau$ for two random walkers moving on a finite $2$-dimensional comb and started at the same point on the backbone to first meet. The comb has a backbone of size $L$ and side chains of length $L$ each, in such a way that overall the comb counts $L(L+1)$ nodes. Data points (bullets) are obtained via numerical simulations (averaged over $10^5$ realizations) and are fitted (solid line) according to the theoretical predictions (\ref{eq:theopred}). Upper inset: In order to check the goodness of the theoretical prediction we plotted the ratio between the experimental value from simulations and the expected value from the analytical estimate. This is done for the analytical prediction given by (\ref{eq:theopred}) and given by a purely power-law (in this case the best-fit exponent $\alpha$ is $\alpha=2.754$). In both cases the ratio is approximately $1$, with fluctuations which are less broadened for the former. Lower inset: We plotted the ratio between the experimental value from simulations and the expected value from the analytical estimate (\ref{eq:theopredd}) for $d$-dimensional combs ($d=2$, $d=3$, $d=4$, as shown by the legend). As expected the ratio fluctuates around $1$.}
\label{mean time}
\end{figure}

During the approaching regime the diffusion along the teeth is, in first approximation, normal \cite{havlin}, in such a way that the standard deviation of the position $Y$ along the teeth scales as $\sqrt{\langle Y^{2} \rangle} \sim \sqrt{t}$, then after time $\tau_1 \sim L^2$, we expect $\sqrt{\langle Y^{2} \rangle} \sim   L$. Being the standard deviation of the same order of the tooth size, we can consider that, when the walker A enters the tooth already occupied by B (referring again to Fig.~\ref{two particles}), the probability distribution of $Y$ is uniform along the tooth. 
The encounter probability $a(Y)$  before A leaves the tooth, neglecting finite size effects \footnote{The encounter probability scales as $a(Y) \sim 1/Y$ only for infinite teeth. The finite size of the teeth yield to a larger encounter probability since the position of B along the common tooth will be biased toward the backbone due to the reflecting boundary conditions at the end nodes.},  is $a(Y)\sim 1/Y$ (see Sec.~\ref{sec:comb}). Therefore, the probability $\mathcal{P}$ that two walkers, sharing the same tooth, meet before one of the two leaves the tooth can be estimated as the average of $a(Y)$ over all positions $Y$:
\begin{equation}\label{eq:encounter_finite}
\mathcal{P}(L) \sim \int_{1}^{L}\frac{1}{L}\frac{1}{Y}dY\sim\frac{\log(L)}{L}.
\end{equation}
This quantity is decreasing with $L$ and, consistently, if the teeth are very long, one of the two walkers can be so far along the tooth that they are unlikely to meet each other (as we have seen for the infinite case). 
%From Eq. \ref{eq:encounter_finite} the average number of returns is proportional to $\frac{L}{\log(L)}$. 
The inverse of the quantity in Eq.~\ref{eq:encounter_finite} can be taken as an estimate for the average number of times that the walkers are found in the same tooth ``trying'' to meet, before they actually succeed in meeting.
Summarising, the mean encounter time $\tau$ is due to the mean time ($\sim L^2$) for sharing the same tooth multiplied by the mean number of times $\mathcal{P}^{-1}$ ($\sim L/\log(L)$) needed for the encounter to effectively occur, namely, $\tau$ scales as:
\begin{equation} \label{eq:theopred}
\tau \sim \frac{L^{3}}{\log(L)}.
\end{equation}
This analytical result is in a very good agreement with numerical simulations as shown in Fig.~\ref{mean time}. 

% In the numerical simulation, the two walker start from the same position on the backbone. They move on a finite comb with size $L$ of the backbone and of the teeth and with reflecting boundary conditions. The simulation stops only when the two walkers find themselves on the same position at a certain time $\tau$. This simulation is performed $10^5$ times for every size $L$ and the average value of encounters times $\tau$ is shown in Fig. \ref{mean time}.

%We have found a little different mean time respect to \cite{agliari blumen}, maybe because sometimes it is difficult to discriminate between a logarithmic and a small power dependence.
In \cite{agliari_blumen} it was also shown that the mean encounter time for two walkers starting with a distance $L/2$ in a $2$-dimensional square comb is $L^{3}$. Such a scaling can be understood within our picture as well. In fact, the
number of steps required to cover a distance $\sim L/2$ on a chain
of length $L$ with the waiting time $\sim L$ scales as $\sim L^{3}$ \cite{redner}.
 Once they share the same tooth, they wait
a time $\sim L^{3}/\log(L)$ to encounter, as just explained before;
but $L^{3}/\log(L)$ is negligible respect to $L^{3}$, which turns out to be the leading
term. 

Finally, for higher-dimensional combs, analogous arguments suggest that the mean encounter time scales as 
\begin{equation} \label{eq:theopredd}
\tau \sim \frac{L^{d+1}}{\log(L)},
\end{equation}
and this is successfully checked in Fig.~\ref{mean time}.

%FORSE OCCORRE SPECIFICARE IL PUNTO DI PARTENZA\\
%TRA L'ALTRO ANCHE NEL CASO 2D NON LO SPECIFICHIAMO MENTRE POI CI VIENE FUORI CHE E' IMPORTANTE PERCHE' CAMBIA LO SCALING di $\tau$!
%
%IL PUNTO DI PARTENZA DEVE ESSERE LO STESSO SITO O CMQ UNA DISTANZA CHE NON SCALA CON L, ESEMPIO 2 O 4,...

\subsection{Two random walkers on bundled structures}\label{sec:bundled}

As explained in Sec.~\ref{sec:comb}, the two fundamental terms to be compared for the two-particle transience are the meeting probability when the particles are both in the same tooth (i.e., the exponent $\alpha$) and the distances travelled by the two particles along the teeth before meeting again in the same tooth (i.e., the exponent $\mu$). It is natural to ask how these parameters vary as the base and the fibre of the underlying comb are modified.

%IDEA:
%
%Nel caso a una particella  c'è monotonicità tra la dimensione spettrale e il passaggio da ricorrente a transiente: più aumenta la dimensione spettrale, più il ritorno all'origine è lento, fino a quando per $\tilde{d}>2$  il ritorno all'origine non è più certo.
% 
%Nel caso a due particelle la situazione non è più la stessa, infatti: 
%
%retta= 2-part. recurrent
%
%pettine= 2-part. transient
%
%piano= 2-part. recurrent
%
%Se invece prendiamo le strutture ramificate e facciamo variare la dimensione spettrale della base ritroviamo un altra forma di monotonicità: se aumenta la dimensione spettrale della base allore $\psi(t)$ è più sbrodolata e cosi anche la $h(Y)$ in modo tale che l'incontro tra le particelle è sempre più raro. Allo stesso tempo, se cerchiamo un lower bound per la transienza a due particelle, dobbiamo diminuire la dimensione spettrale delle fibre, ad esempio considerando denti con lunghezza random power law $y^{-\gamma}$. Esisterà un valore $\gamma$ di transizione tra 2 particle transience e two particle recurrence.

Let us consider a generic branched structure (see Fig. \ref{fig:comb}) with base ${\mathcal{B}}$ and fiber ${\mathcal{F}}$ characterized by spectral dimension $\tilde{d}_{\mathcal{B}}$ and $\tilde{d}_{\mathcal{F}}$, respectively. Analogously to the case of the simple comb, while exploring the structure, one of the two walkers will eventually enter the fiber already occupied by the other walker. This phenomenon can be described in terms of the probability $\psi(t)$  for the walkers to first share the same fiber at time $t$, and of the probability $h(Y)$ that the walkers display a distance $Y$ along the fiber.
In general we expect that, by increasing the spectral dimension of the base, the time taken by the two walkers to be in the same fiber gets more broadly distributed. %Indeed, considering the case of fibers given by linear chains, also the jump distribution of relative distance along $y$ becomes more broadly distributed, in fact it depends on $\psi(\tilde {t})$ (see Eq.~\ref{eq: salti h(x)}).
If we take as fibers linear chains, we can see immediately that a broader $\psi(t)$ also implies a broader distribution for the relative distance $h(Y)$ (see Eq.~\ref{eq: salti h(x)}).
Moreover, the probability of encounter along a tooth remains $a(Y)\sim 1/Y$, consequently also in these cases the two-particle transience is ensured (see Sec.~\ref{sec:comb} and the condition of the paper \cite{CASC-PRE2015}) as proven in \cite{Krishnapur-ECP2004,Cassi_Campari_2012}.

However, natural structures often exhibit inhomogeneous teeth. Therefore, more realistic models should include a probability distribution $\chi(L)$ for the teeth length $L$. When the average length of the teeth is finite, the comb can be effectively thought of as a line, in fact diffusion along the backbone is normal \cite{havlin} and the spectral dimension is $\tilde{d}=1$.
% In this case the probability of being in the same tooth (for the two walker) has the same behavior as the probability of returning to the origin for a single walker in a line: $\psi(t) \sim t^{-3/2}$ (see eq.\ref{eq:Fcomb} and eq. \ref{eq:F(t)}). 
%
Conversely, when the average length diverges, diffusion along the backbone becomes anomalous \cite{havlin}. Therefore, we expect that, depending on the distribution $\chi(L)$, the walkers meet with certainty or not.

In particular, in wedge combs, namely structures where the length of the teeth is given by a deterministic function $f(x)=x^{\delta}$ of the position $x$ along the backbone, the two-particle transience appears if and only if $\delta>1$ \cite{wedge_comb}. 
This is consistent with our framework since when $\delta>1$, the average length of the teeth is infinite.
%Indeed in this case (as for random combs) the average length of the teeth is infinite. 
%Therefore, the results for a random distribution of teeth length can be seen as a generalization to the disordered case of the (ordered) wedge combs.  

%LASCIO QUESTI COMMENTI COME HINTS TO BE POSSIBLY INVESTIGATED IN FUTURE... EVENTUALMENTE POSSIAMO INSERIRLI NELLE CONCLUSIONI COME OUTLOOKS\\
%UNA LUNGHEZZA MEDIA INFINITA PER LE SIDE CHAIN E' SUFFICIENTE A DEDURRE LA TRANSIENZA A DUE PARTICELLE?\\
%POSSIAMO DEDURRE QUALCOSA PER LO HIERARCHICAL COMB? (PIU' CHE ALTRO PERCHE' ANCHE QUELLO E' UN MODELLO PIUTTOSTO FAMOSO E STUDIATO) non conoscendo bene il limite termodinamico dell'hirarchical comb, non so bene cosa dire...
%
%
%Ora la questione dei denti random: anche qui bisogna dividere il moto lungo x da quello lungo y:
%
%diffusione lungo x è nota: vedi libro di havlin
%
%diffusione lungo y: come è la diffusione su denti di lunghezza random? la distribuzione di probabilità $p(y,t)$ della posizione y  su segmenti di taglia finita L è nota in serie di Fourier. Poi va mediata sulla distribuzione di L, cioè sulla power law. Forse si potrebbe anche trovare in questo modo la $h(Y)$ ma poi cmq ci sarebbe da trovare la $a(y)$. La forma della $a(y)$ mi è nota  solo nel caso di rette infinite...ma nel caso con L finito non so cosa succeda. Forse anche nel caso finito, la $a(y)$ è nota ma credo che diventi una cosa troppo lunga... quindi lascerei perdere. Cosa ne pensi? Qualche idea?
%
%se sapessi come trattare la diffusione lungo y, allora avrei la soluzione!

\section{Checking the robustness via topological perturbations \label{sec:ponti}}

Comb graphs are two-particle transient, while two-dimensional Euclidean lattices are two-particle recurrent. Hence, by inserting bridges between the teeth of the comb, the latter becomes more and more
similar to a two-dimensional lattice and will eventually loose the two-particle transience. Here we want to investigate numerically such a transition: we start from a simple comb of linear size $L$ (i.e., the length of the backbone and of the teeth is $L$) and we insert $2 \,L^{\alpha}$ ($\alpha \in [0,2]$) edges between couples of nodes belonging to adjacent teeth and lying at the same height with respect to the backbone. Otherwise stated, if we imagine the comb embedded in a two-dimensional lattice, we are inserting $2 L^{\alpha}$ horizontal links of unitary length. The bridges inserted are scattered randomly among the $2L^2$ available slots. 
\newline
Once the structure is generated, we perform Monte Carlo simulations where the two walkers start from the same position on the backbone and are made run until they meet or until the number of time steps is larger than $L$; the latter condition ensures that the walkers have not reached the borders of the comb, and therefore that our results are not biased by finite-size effects.
\newline
For a given realization of the underlying structure this process is repeated $10^3$ times in order to get a good statistics over the possible paths of the two walkers; the encounter probability, referred to as $P_{\textrm{enc}}(\alpha,t)$, is then estimated as the ratio between the number of encounters occurred by the time $t$ divided by the total number of simulated paths. A further average over $10^2$ different realizations of the underlying structure is then accomplished and we get the mean encounter probability $\bar{P}_{\textrm{enc}}(\alpha,t)$.
%The exact position on the backbone is irrelevant because we chose the periodic boundary condition on the x-axis.
%
%3 sizes of the graphs: $L=2^{11},\,2^{13}, \,2^{15}$. 
%
\newline
The mean encounter probability $\bar{P}_{\textrm{enc}}(\alpha,t)$ is then fitted with a function $y(t)=P_{\textrm{enc}}^{\infty}(\alpha) -g(t)$, with $g(t) \to 0$ as $t\to \infty$, in such a way that the fit coefficient $P_{\textrm{enc}}^{\infty}(\alpha)$, corresponding to the asymptotic value of $\bar{P}_{\textrm{enc}}(\alpha,t)$, provides our estimate for the probability of eventually meeting: $P_{\textrm{enc}}^{\infty}(\alpha)$ equals $1$ if the underlying structure is two-particle recurrent, while it is strictly smaller than $1$ if the underlying structure is two-particle transient.

\begin{figure*}
\includegraphics[width=1\linewidth]{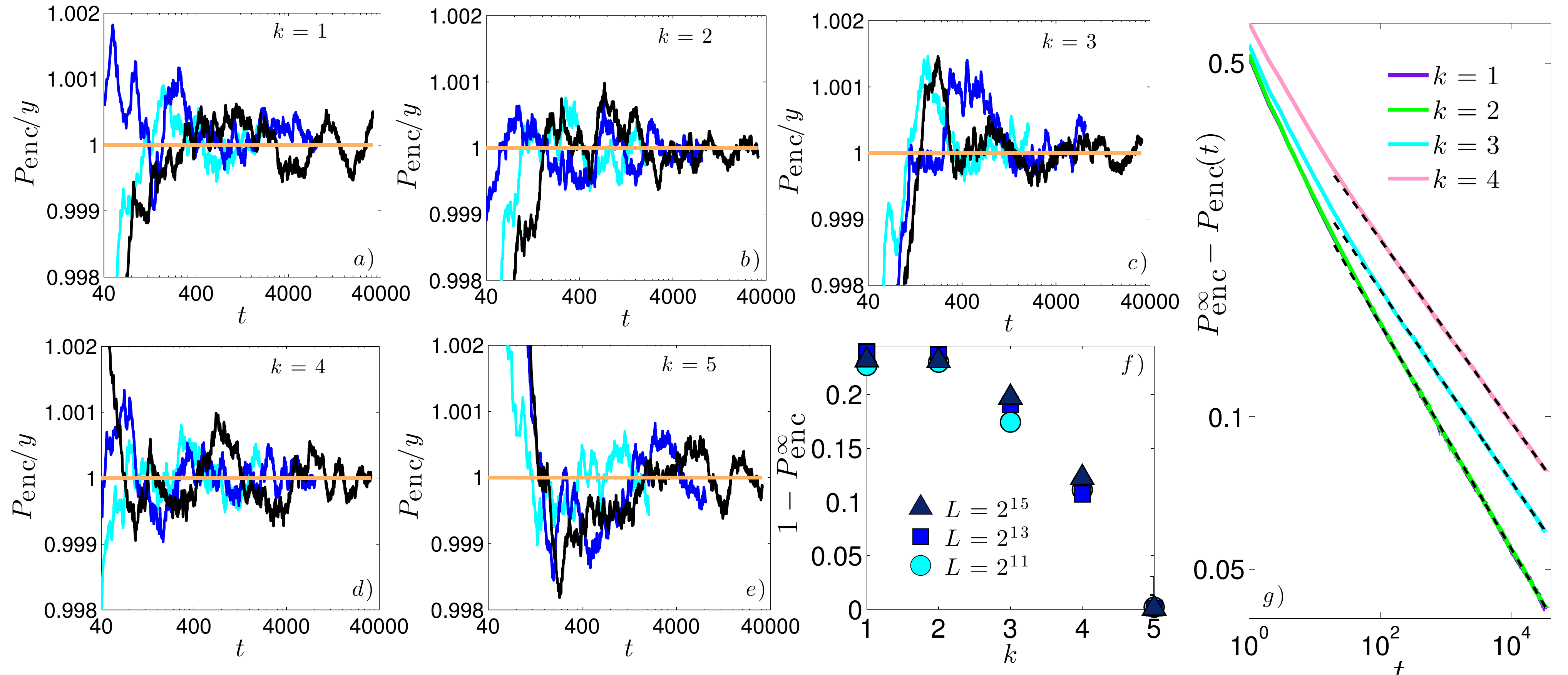}
\caption{(Color on line)
Panels $a$-$e$: Ratio between the numerical estimate of the encounter probability $\bar{P}_{\textrm{enc}}$ and the value provided by the fitting function; each panel corresponds to a different choice of $k$ (i.e., a different choice of $\alpha$). For $k=1,2,3,4$, the fitting function is $y(t) = P_{\textrm{enc}}^{\infty}(\alpha) -a t^{-b}$ while for $k=5$ it is $y_1(t) = P_{\textrm{enc}}^{\infty}(\alpha)-a t^{-b}-c/\log(t)$. 
In each panel we compare results for three different system size: $L=2^{11}$ (bright blue), $L=2^{13}$ (blue), and $L=2^{15}$ (black). 
The best-fit coefficient $P_{\textrm{enc}}^{\infty}(\alpha)$ is used in panel $f$ to show how the probability of never meeting varies as the number of links inserted is progressively increased. For the cases analyzed, the encounter is certain only for $\alpha=2$. 
The best-fit coefficient $P_{\textrm{enc}}^{\infty}(\alpha)$ is also used in panel $g$, where we plot, in a log-log scale,  the difference $P_{\textrm{enc}}^{\infty}(\alpha) - \bar{P}_{\textrm{enc}}(\alpha,t)$ pertaining to the cases $k=1,...,4$. The dashed black lines have slope $-b$. The linear outline versus time corroborates the expected power-law behaviour for the related encounter probabilities. }
\label{simu}
\end{figure*}

More precisely, in our numerical experiments we addressed 5 different cases, labelled by $k$: $\alpha=0$ ($k=1$), $\alpha=1$ ($k=2$),  $\alpha=1.5$ ($k=3$), $\alpha=1.8$ ($k=4$) and $\alpha=2$ ($k=5$); notice that the case $k=1$ roughly corresponds to a simple comb, while the case $k=5$ corresponds to the square lattice.
\newline
For all the cases analyzed, but the case $k=5$ (i.e., the square lattice), a power law $y(t) = P_{\textrm{enc}}^{\infty}(\alpha) -a t^{-b}$  provides a successful description for the temporal evolution of $\bar{P}_{\textrm{enc}}(\alpha,t)$, in fact, this function is the typical time saturation law for this kind of processes \cite{havlin}. For the two-dimensional Euclidean lattice the situation is different and the asymptotic behaviour of the encounter probability is known to depend 
logarithmically on time.  Indeed, thanks to the homogeneity of the lattice, the encounter probability between two random walkers is   asymptotically equivalent to the probability of return to the origin for the single walker \cite{Cassi_Campari_2012} which is $\sim 1- c/ \log(t)$ \cite{dvoretzky1951some}. Therefore, in order to fit $\bar{P}_{\textrm{enc}}(2,t)$, we need to add a logarithmic term to the fitting curve: $y_1(t) = P_{\textrm{enc}}^{\infty}(\alpha)-a t^{-b}-c/\log(t) \approx P_{\textrm{enc}}^{\infty}(\alpha)-c/\log(t)$ (note that the power-law term is asymptotically negligible for $t \to\infty$).  
The ratio $\bar{P}_{\textrm{enc}}/y$ for all the cases analyzed is reported in Fig.~\ref{simu} (panels $a$-$e$). 
\\
Now, a few remarks are in order. In general fits are very good with a discrepancy smaller than $1 \%$, at least for relatively long times. This is checked for three different sizes, namely $L=2^{11},\,2^{13}, \,2^{15}$. When $\alpha<2$ (i.e., $k<5$), both fitting functions $y(t)$ and $y_1(t)$ provides very good fits with $1 - R^2 <10^{-3}$, and the related estimates for $P_{\textrm{enc}}^{\infty}(\alpha)$ (namely the constant term of the fitting function) are, within the error ($\sim 5 \%$), comparable, hence the former is preferred since a smaller number of parameters is involved.  
The reliability of the power-law fit for $\alpha<2$ is further inspected by plotting $P_{\textrm{enc}}^{\infty}(\alpha) - \bar{P}_{\textrm{enc}}(\alpha,t)$ in a log-log scale and checking that the outline is linear with slope corresponding to the related fit coefficient $b$ (see Fig.~\ref{simu}, panel $g$). 

These analysis corroborate the reliability for our estimates of $P_{\textrm{enc}}^{\infty}(\alpha)$, which are summarized in Fig.~\ref{simu}, panel $f$. Again, several sizes are compared and overall the estimates seem to be robust.
In particular, as long as $\alpha = 2$, we get that $P_{\textrm{enc}}^{\infty}(\alpha)=1$, meaning that  encounter is certain, while when $\alpha <2$, we get that $P_{\textrm{enc}}^{\infty}(\alpha)<1$, meaning that there is a finite probability that the two walkers will never meet.

We note that in the thermodynamic limit, the number of additional links is infinite for $\alpha>0$, then the two-particle transience is expected to be preserved even under the introduction of an infinite number of loops. On the other hand, the density $\rho = 2 L^{\alpha}/L^2$ of additional links is zero for $\alpha<2$, hence suggesting that a sub-linear (in the volume) number of additional links is not sufficient to break the two-particle transience.

%The topology that we considered represents the number of loops in the network:
%Comb (no further loops), "L", where the number of loops in the network scales linearly with the length of the backbone, "$L^{1.5}$", "$L^{1.8}$" where the number of loop scales with a specific power of L, and "$L^2$", where the number of loops is maximized.
%
%In the simulations, first we 'build' the graphs: we start from a comb graph and we add "N" bridges between first neighbour teeth. %A point can be added only between two points that are s if we embed the graph in the Euclidean space. 
%We considere a comb-base where the backbone's size is L and the teeth are of size L too. 
%In total there are $2 L^2$ possible loops that can be added. The number of loops that we added for each topology is respectively: 0, $2 L$, $2 L^{1.5}$, $2 L^{1.8}$, $2 L^2$.

%We let evolve the particle for a time $T=2^L$, in order to be sure that our results aren't affected from the finite-boundary.

\section{Mapping the two-particle problem into a one-particle problem \label{sec:The Mapping}}

In general, the encounter of two walkers on a given structure can be mapped into a one-particle problem, where the (first) encounter  probability is rewritten as the probability to (first) reach a given set of sites. In this mapping we reduce the number of particles involved, but we pay a price in terms of topological complexity since the one-particle problem turns out to be embedded in a structure which is typically tougher than the original one.
Anyhow, the mapping can still be convenient in order to solve or to deepen the problem considered.

Let us focus on two walkers moving in a two-dimensional comb and notice that the temporal evolution of their relative distances (i.e., $\Delta_{x}$ and $\Delta_{y}$) and of their (non normalized\footnote{Notice that, for mathematical convenience, in this mapping we adopt as variables $\tilde{X}_{\textrm{cm}} \equiv (X_1 + X_2)$ and $\tilde{Y}_{\textrm{cm}} \equiv (Y_1 + Y_2)$. These correspond to the instantaneous position of the center of mass of the system, namely $(X_{\textrm{cm}} , Y_{\textrm{cm}}) \equiv ((X_1 + X_2)/2,(Y_1 + Y_2)/2)$, apart for a factor $2$. }) centers of mass (i.e., $\tilde{Y}_{\textrm{cm}} \equiv Y_1 + Y_2$ and $\tilde{X}_{\textrm{cm}} \equiv X_1 + X_2$) completely characterizes the system.
Actually, by exploiting the translational invariance along the backbone, the evolution of the variables $(\Delta_x,\,\Delta_y,\, \tilde{Y}_{\textrm{cm}})$ is completely independent of the value of $\tilde{X}_{\textrm{cm}}$, which is therefore unnecessary in describing the encounter between the walkers and can be neglected in building the mapping. 
Now, the set of variables $(\Delta_x,\,\Delta_y,\, \tilde{Y}_{\textrm{cm}})$ effectively describes a single walker in a proper structure, referred to as $\mathcal{M}$, which is schematically shown in Fig.~\ref{mapping-1} (see Appendix \ref{mapping_appendix} for more details on the construction of $\mathcal{M}$).  

\begin{figure}[tb]
\noindent \begin{centering}
\includegraphics[width=0.40\textwidth]{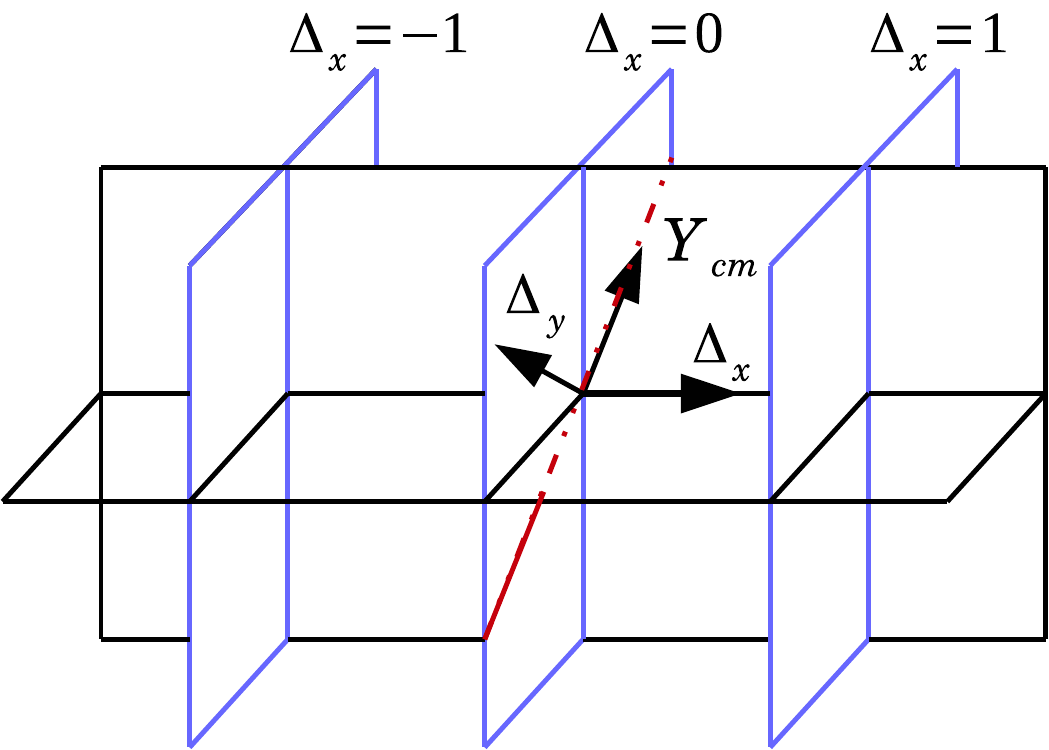}
\includegraphics[width=0.20\textwidth]{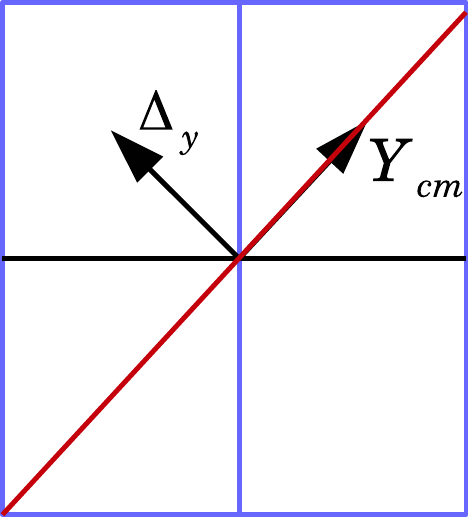}
\par\end{centering}
\caption{ (Color online) Upper panel: the two-particle problem on a $2$-dimensional comb can be mapped into a one-particle comb embedded in a structure as the one shown here. Every point of this structure is univocally associated to a triplet $(\Delta_x,\,\Delta_y,\,Y_{\textrm{cm}})$ and the encounter between the two walkers on the comb corresponds to the single particle being in any point of the straight line $\Delta_x=\Delta_y=0$ denoted in red. Lower panel: the plane $\Delta_x =0$ (called the 'plane of encounters') is shown alone to highlight the encounter line along $Y_{\textrm{cm}}$.}
\label{mapping-1}
\end{figure}

The starting point of the walker in $\mathcal{M}$ is the origin of axes (where $\Delta_x=\Delta_y=\tilde{Y}_{\textrm{cm}}=0$) which corresponds to let the two particles start at the same point in the backbone of the comb. We also outline a set of points (the line along $Y_{\textrm{cm}}$ in Fig.~\ref{mapping-1}) referred to as the \emph{encounter line}. In fact, the random walker reaching any of these points in $\mathcal{M}$ corresponds to the encounter (i.e., zero relative distances: $\Delta_x=\Delta_y=0$) of the two random walkers in the original comb. 

Now, in this mapping we can recover all the properties discussed in Secs.~\ref{sec:comb} and \ref{sec:further}.
For example, every time the walker in $\mathcal{M}$ returns to the plane $\Delta_x =0$ (namely, when the two particles in the comb return in the same tooth) its coordinate $\Delta_y$ is taken from a probability distribution $h(\Delta_y)\sim \Delta_y^{-3/2}$ as in Eq. \ref{eq: salti h(x)},  where we negelcted the absolute value to increase the readability. The probability $a(\Delta_y)$ to visit the encounter line before leaving the plane $\Delta_x =0$ scales as $\sim 1/ \Delta_y$ \cite{redner}. Therefore, recalling again that for a walker with jump distribution $h(\xi) \sim \xi^{-\mu -1}$ in the presence of traps with distribution scaling as $a(\xi) \sim \xi^{-\alpha}$, the overall absorption is not certain as long as $\mu<\alpha$ \cite{CASC-PRE2015}, we get that the walker on $\mathcal{M}$ has a finite probability not to meet the encounter line. This basically equals to state the two particle transience of the comb.\\

The situation where the particles have different %velocities $v_1 , v_2$, with $v_1/v_2 \neq 0$ and $v_1/v_2 \neq \infty$ (namely both particles are moving), see Sec.~\ref{sec:velocities},
diffusivities
 can as well be addressed: the change of the diffusivities generates a rotation of the line of the encounter in $\mathcal{M}$ which still does not change the asymptotic behaviour of  $a(\Delta_y)$ and of $h(\Delta_y)$.

%\%\% QUI C'È LA PARTE CON LE DUE DIFFUSIVITÀ CHE VA ESPRESSA IN TERMINI SIMILI ALLA SOTTOSEZIONE PRECEDENTE SUL CASO A DUE DIFFERENTI DIFFUSIVITÀ

Finally, the mapping introduced allows getting the same results in the finite-size problem, where the planes of Fig.~\ref{mapping-1} are finite (see Sec.~\ref{sec:finite_comb}) and in other bundled structures, for example $d$ dimensional combs, brushes, bundled fractals (see Sec.~\ref{sec:bundled}). In fact, when the fibre of the bundled structures are lines, the mapping is characterized by an infinite number of planes $(\Delta_y,\,\tilde{Y}_{\textrm{cm}})$ like in Fig.~\ref{mapping-1}, even if differently linked each other.

\section{Conclusions} \label{sec:concl}
In this work we developed an effective framework for the analytical investigation of the two-particle problem on comb-like structures. In fact, in such inhomogeneous architectures the two-particle problem (i.e., the problem of finding the probability that two random walks will eventually meet) can differ qualitatively from the one-particle problem (i.e., the problem of finding the probability that a random walk will eventually reach a given site) and having tools for deepening this phenomenon is of crucial importance not only from a theoretical perspective, but also from an experimental one (e.g., to unveil whether the reaction is favoured by either a fixed or a mobile target). 
 
After having outlined our analytical framework meant for general branched structures, we explicitly studied some specific examples.
In particular, we recovered that in simple two-dimensional combs, in the limit of infinite size, there is a finite probability that two walkers will never meet, no matter their initial positions (see \cite{Krishnapur-ECP2004,Chen-JProb2011,Cassi_Campari_2012} for a rigorous proof). This feature is also referred to as ``two-particle transience''.
Moreover, we derived the probability that the encounter (if any) will occur in the backbone or in a teeth; 
remarkably, we evidenced that, asymptotically, the average number of encounters in the backbone is finite, while the average number in teeth is infinite. 
\newline
We also showed that the two-particle transience is robust with respect to changes in the diffusitivity of the walkers (provided that both walkers are effectively moving) and with respect to changes in the topology of the base.
\newline
Our framework can also account for finite structures and we obtained a description for the slowing down of the two-particle reaction in finite, two-dimensional combs, already evidenced numerically in \cite{agliari_blumen}.
\newline
Finally, in order to further investigate the robustness of the two-particle transience we studied numerically the asymptotic first-encounter probability for two random walkers set in a two-dimensional combs where short-cut among teeth are progressively inserted. Interestingly, we found that, as long as the number of links inserted scales sub-linearly with the volume, the two-particle transience is preserved.

This work can also be a starting point to explore many-particle phenomena on branched structures, which, as shown in \cite{autocatalitica}, may deviate qualitatively from mean-field predictions. For instance, in reactions such as the autocatalytic, the coalescence or the annihilation, we expect that the evolution of the species concentration will mirror the two-particle transience with non trivial outcomes.

We also believe that many physical applications can take advantage of the two-particle transience. If, for example, one needs to slow down a reaction between different elements taking place on a Euclidean lattice, it could be useful to use geometries-based strategies and properly cut edges (hence moving toward a comb-like architecture) rather than add edges (as one could naively imagine).

%Moreover the mapping introduced here can be useful to approach several extensions of the two-particle problem. For instance, general bundled structures can be easily addressed in these terms, and the encounter time can be derived even at a finite size scale. Also, the mapping can represent even an interesting bridge between the one particle problem and the two-particle problem from the opposite perspective: one could recast a one particle problem embedded in a inhomogeneous structure into a two particles problem embedded in a easier structure. 

\appendix
\section{Encounter probability}\label{app}

\begin{figure}[h]
\noindent \begin{centering}
\includegraphics[width=0.35\textwidth]{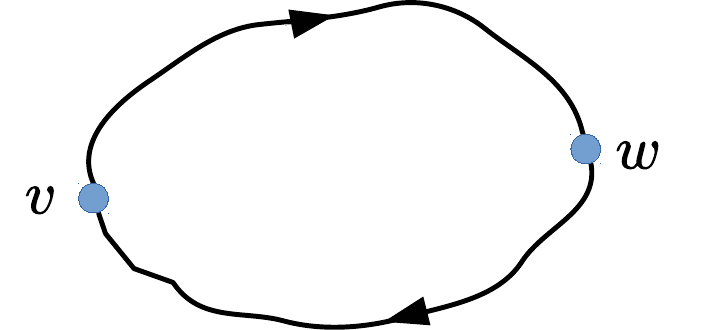}
\par\end{centering}
\caption{ (Color online) The return to the origin $v$ for a walker, passing  through $w$.}
\label{ciclo}
\end{figure}

The encounter between two random walkers can display deeply different properties according to whether the underlying structure is either homogeneous or inhomogeneous, yet there exist general results valid for every graph with finite degree.  In these graphs we  show a deep relation between the two-particle encounter probability and the one-particle probability of return to the origin, that is, we will show that they have the same asymptotic behavior. We start analysing the motion of the single walker assuming the Markov property\footnote{This assumption is crucial for the following derivation, while, of course, it may not hold in realistic applications.}. When a random walker  starts from a given site $v$ and then, eventually, it is back to the origin $v$, it visits an arbitrary site, say $w$, in such a way that we can decompose the cycle into the path from $v$ to $w$ and the path from $w$ to $v$ (see Fig. \ref{ciclo}).

The probability $P_{vw}(t)$ of reaching $w$, being started at $v$, is equal to the probability $P_{wv}(t)$ of reaching $v$ starting from $w$, except for a factor $z_w/z_v$ accounting for the (possibly different) coordination number $z$ of the starting and final sites.
To calculate the probability of return to $v$, we have to sum over all possible visitable sites $w$:
\begin{equation}
P_{vv}(2t)=\sum_{w \in V}P_{vw}(t)P_{wv}(t)=\sum_{w \in V}[P_{vw}(t)]^{2}\frac{z_{w}}{z_{v}},\label{eq: andata ritorno}
\end{equation}
where the first equality stems from the Chapman Kolmogorov equation and $V$ is the set of sites making up the underlying graph.

\begin{figure}[h]
\noindent \begin{centering}
\includegraphics[width=0.35\textwidth]{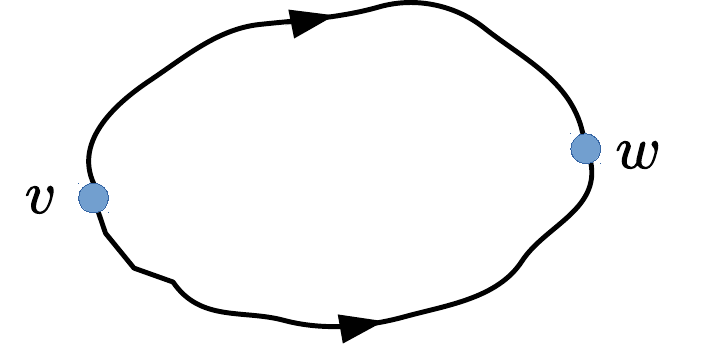}
\par\end{centering}
\caption{ (Color online) Two walkers start form $v$ and collide in $w$.}
\label{due-particelle}
\end{figure}

Now, let us consider the case of two walkers and let us denote with $\boldsymbol{P}_{(vv)}(t)$ the probability that, being both started from $v$, they will meet in any site $w \in V$. One can see that $\boldsymbol{P}_{(vv)}(t)$ is related to the probability $P_{vw}(t)$ of the single particle (see Fig. \ref{due-particelle}) as
\begin{equation}
\boldsymbol{P}_{(vv)}(t)=\sum_{w\in V}P_{vw}(t) P_{vw}(t)=\sum_{w\in V}[P_{vw}(t)]^{2}.\label{eq: una e due particelle}
\end{equation}

When the node degrees are everywhere finite and bounded with maximum $z_{max}$ and minimum $z_{min}$, we get the following lower and upper bound (see Eq. (\ref{eq: andata ritorno})):
\begin{equation}
\frac{z_{min}}{z_{max}}\sum_{w} \left[ P_{vw}(t) \right]^{2}\leq P_{vv}(2t)\leq \frac{z_{max}}{z_{min}}\sum_{w} \left[ P_{vw}(t) \right]^{2},
\end{equation}
then, using Eq. (\ref{eq: una e due particelle}), we get
\begin{equation}
\frac{z_{min}}{z_{max}}\boldsymbol{P}_{(vv)}(t)\leq P_{vv}(2t)\leq \frac{z_{max}}{z_{min}}\boldsymbol{P}_{(vv)}(t).\label{eq: lower e upper}
\end{equation}
From Eq.~\ref{eq: lower e upper} we deduce that $\boldsymbol{P}_{(vv)}(t)$ and $P_{vv}(2t)$ have the same asymptotic trend, that is, in general the probability of encounter for two walkers has the same trend as the probability of return to the origin. This is a very strong relation between the one- and the two-particle problem. Anyway, this result is not in contrast with the splitting between one-particle recurrence and two-particle transience occurring in highly inhomogeneous graphs (e.g. combs), as this feature concerns non Markovian quantities, such as first passage and first encounter probabilities.  An important question arises: what happens in infinite degree graphs or in networks  with infinite average degree (e.g., with degree distribution $P(k)\sim k^{-\gamma}$, $\gamma\leq 2$)?

\section{Some details about the construction  of the mapping}\label{mapping_appendix}

In order to map the two-particle problem into a
one-particle problem, one could study the evolution of the coordinates of the two walkers $X_{1},\, X_{2},\, Y_{1},\, Y_{2}$ in a four dimensional space. However, this does
not really simplify the problem because the resulting new underlying
topology is by far not trivial. 

A better approach is focusing on the (non-normalized) center of mass, corresponding to the coordinates
$\tilde{X}_{\textrm{cm}},\, \tilde{Y}_{\textrm{cm}}$, and on the relative distances along $x$ and $y$, referred to as
$\Delta_{x}$ and $\Delta_{y}$, respectively. Now, we can take advantage of the symmetry along
$x$ displayed by the comb: as the motion of the particles and
their encounters are independent of $X_{\textrm{cm}}$, this variable can be neglected.
The meeting corresponds to reaching the line with $\Delta_{x}=\Delta_{y}=0$.

\begin{figure}[h]
\noindent \begin{centering}
\includegraphics[width=0.35\textwidth]{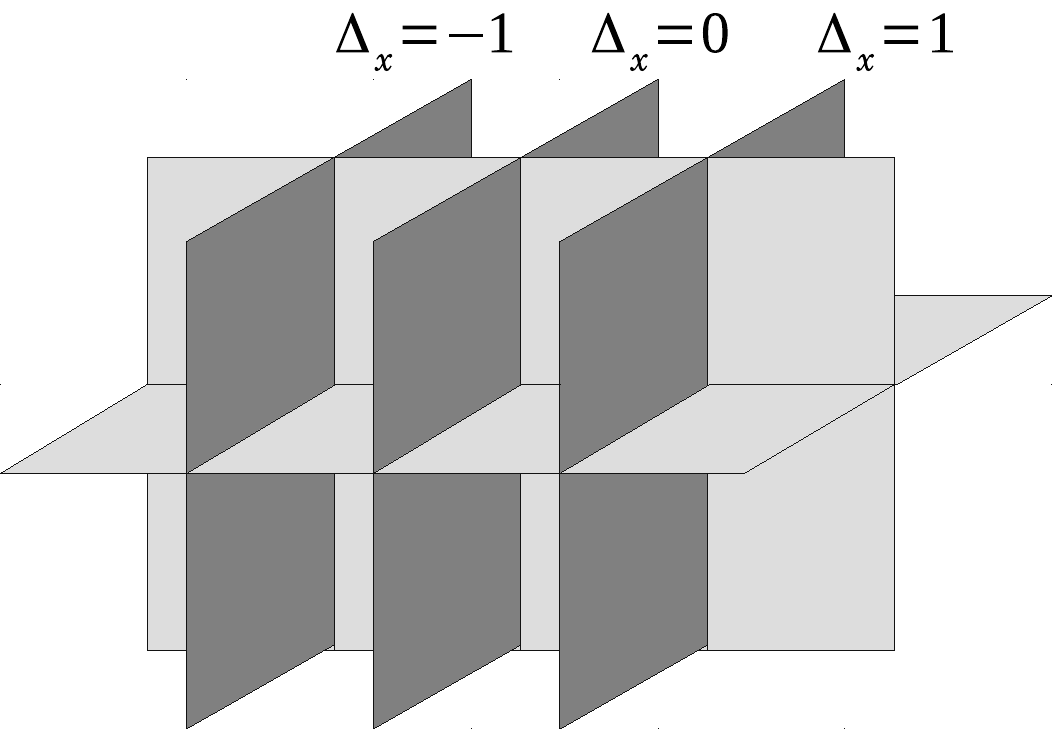}
\par\end{centering}
\caption{ (Color online) In this schematic representation of $\mathcal{M}$, we distinguish the ``pages'' (in dark color) and the ``bookbindings'' (in bright color). }
\label{mapping_colori}
\end{figure}

In this three-dimensional space (with coordinate axes $Y_{\textrm{cm}},\, \Delta_{x},\, \Delta_{y}$),
we construct the graph $\mathcal{M}$ (sketched in Fig.~\ref{mapping-1}) by considering
the whole set of possible motions on the comb:

\begin{enumerate}

\item \emph{Both particles on the teeth}: the possible motions are $4$,
each with probability \textonequarter{}: \textdownarrow{} \textuparrow{},
\textuparrow{} \textdownarrow{}, \textdownarrow{} \textdownarrow{},
\textuparrow{} \textuparrow{}. This is the case studied by Polya \cite{Polya} who showed that the random motion of two walkers on a line can be mapped into the random walk of a particle in  a plane, where the previous four possible movements are mapped into the $4$ directions \textdownarrow{}, \textuparrow{}, \textrightarrow{}, \textleftarrow{} on the plane. Since in this situation the two walkers can not 
modify their relative distance
along $x$, in the mapping we will consider infinite parallel planes, each corresponding to a different value of $\Delta_x$. We call ``page'' each of these infinite planes (shown in dark shadow in Fig.~\ref{mapping_colori}).

\item \emph{At least one of the two particles on the backbone}: in this
case it is possible to vary the relative distance along $x$ between
the two particles so the walker in the mapping is allowed to move
from one plane to another with different $\Delta_{x}$. If at least one of the two particles is
on the backbone, we have $\Delta_{y}=\pm \tilde{Y}_{\textrm{cm}}$, which represent
two planes that intersect each other as well as those introduced in the previous point. 
We call ``bookbinding'' each of these planes (shown in bright shadow in Fig.~\ref{mapping_colori}).
\end{enumerate}

As mentioned in Sec.~\ref{sec:The Mapping}, in this mapping it is convenient to use $\tilde{Y}_{\textrm{cm}}=Y_1+Y_2$, instead of $Y_{\textrm{cm}}=(Y_1+Y_2)/2$ since, in the latter case, the equations for the bookbindings would be $\Delta_{y}=\pm 2 Y_{\textrm{cm}}$ (in fact, a particle on a tooth at distance $Y$ from the backbone and another particle on the backbone display a distance $\Delta_y = Y$, but the center of mass is located in $Y_{\textrm{cm}} = Y/2$), and this would make the notation and the structure itself a bit more complicated. On the other hand, if we use $\tilde{Y}_{\textrm{cm}}$, the equation is just $\Delta_{y}= \pm \tilde{Y}_{\textrm{cm}}$.

\subsection{Simplifying assumptions\label{sub:Simplifying-assumptions}}
The motion on $\mathcal{M}$ is rather complex as it is subject to a number of constraints. 
However, since we are interested in the long time behavior, we can take advantage of the robustness of the asymptotic
properties with respect to local details, and neglect several of them (see e.g., \cite{burioni1}), hence significantly simplifying
the problem. In particular, 
\begin{enumerate}
\item if $\Delta_{x}$ is even, then $\Delta_{y}$ and $\tilde{Y}_{\textrm{cm}}$ are even; if $\Delta_{x}$
is odd, then $\Delta_{y}$ and $\tilde{Y}_{\textrm{cm}}$ are odd; these constraints can
be neglected because they do not modify the topological structure
of the pages;
\item when the walker is in a bookbinding, it can jump to two sites
belonging to the nearest page on one side and to other two sites
belonging to the nearest page on the other side. Here, exploiting the fact that
the local topological details are irrelevant to determine the graph
type (i.e., either recurrent or transient), we will allow the random
walk to jump only to one (instead of two) sites in each nearest page.
Indeed, it is possible to show that recurrence and transience are
left invariant by adding and cutting links satisfying the quasi-isometry
conditions;
\item when both particles are on the backbone ($\Delta_{y}= \tilde{Y}_{\textrm{cm}}=0$), the walker
on the mapping may also go to the two next-nearest neighbour pages.
We will not consider this possibility for the same reason of previous
point;
\item some sites in $\mathcal{M}$ are actually endowed with waiting times, which arise because, if
both particles are moving in the same direction on the backbone, $\Delta_{y},\, \tilde{Y}_{\textrm{cm}},\, \Delta_{x}$
remain unchanged. We can ignore these waiting times because any local bounded
rescaling of the transition probabilities and waiting probabilities associated to links leave
the random walk type unchanged.
\end{enumerate}

\subsection{The plane of the encounters}

The starting point of the random walk defined in $\mathcal{M}$ is
the origin of axes (which corresponds to take the two particles at
the same starting point in the backbone of the comb) and the meeting
between the two particles corresponds to returning to the line of
encounters $Y_{\textrm{cm}}$ (in Fig. \ref{mapping-1}), where the relative
distances are zero.

We therefore focus
only on the plane embedding this line  (see the lower panel of Fig. \ref{mapping-1}), which represents the case when the two walkers on the comb are in the same tooth (since $\Delta_x=0$).

%In this plane the chance to escape into the structure of (\ref{mapping-1})
%is mapped into the possibilities of waiting times and jumps at the
%$X$-axis and $Y$-axis sites, where there are the blue circles:
When the walkers moves in this plane, and occurs to to be in any point of the lines $Y_{\textrm{cm}}=\pm \Delta_y$, it means
that it has the possibility of getting lost in the other pages of
the graph and come back after a time $\tilde{t}$ in another site of the lines $Y_{\textrm{cm}}=\pm \Delta_y$. Therefore when the walker occurs to be in a site of the lines $Y_{\textrm{cm}}=\pm \Delta_y$,
it could jump to another site of these lines with a waiting time $\tilde{t}$.

The length of the jumps has a Gaussian distribution with variance
proportional to the waiting time $\tilde{t}$, because if we look at the
structure of $\mathcal{M}$ (see Fig. \ref{mapping-1}), the walker that
leaves the plane of the returns, whatever path it follows, always
moves in planes. If a walker diffuses in a plane, its probability
distribution evolves as a two-dimensional normal distribution with
variance proportional to the time of the evolution. 

%DA QUI FACCIO IL PARALLELO CON L'INCONTRO SU PETTINE:
When the walker goes on the line of jumps, it comes back into the
plane of encounters after a delay due to the time for the two particles
to share again the same tooth.

Let us now derive the probability distribution $\psi(\tilde{t})$ for the walker in $\mathcal{M}$ 
to first return to the plane of encounters $\Delta_x =0$. We preliminary notice that, given the symmetry of the two ``bookbindings'', we can collapse them\footnote{This procedure is similar to the collapse of the branches in a Bethe lattice, when interested in the return to the root.} and just focus on the motion of a walker on $\mathcal{M}$ where only one of the two ``bookbindings'' is retained, this just implies sub-leading corrections. This simpler structure is the Cartesian product between a line and a comb, in such a way that
the probability distribution $\psi(\tilde{t})$ to first return to the plane $\Delta_x=0$ is completely determined 
by the properties of the comb, on which the probability of return to the origin for the first time scales as $\tilde{t}^{-5/4}$ \cite{bundled,burioni1}.
%
% $\tilde{t}$ has a probability distribution given by the probability distribution
%$F_{0}(t)$ of the first return to the origin for a single particle.
%Being $\tilde{d}_{comb}=3/2$ and $F_{0}(t)\sim t^{min\{\tilde{d}/2-2,-\tilde{d}/2\}}$
%(see \cite{burioni1}).
%
%
The asymptotic distribution therefore reads as 
\begin{equation}
\psi(\tilde{t})\sim\tilde{t}^{-5/4}.\label{eq: waiting time distribution}
\end{equation}

The details that we mentioned before in Appendix \ref{sub:Simplifying-assumptions}
have no effect on the asymptotic probability distribution of
$\tilde{t}$ \cite{burioni1}, so we can safely neglect them since we are
only interested in the asymptotic properties.  Obviously, the variable $\tilde{t}$ corresponds
to the variable $t$ in Eq.~\ref{eq:Fcomb}.

As a first  check, we simulated a random walk in $\mathcal{M}$ and verified that its diffusion properties are consistent with the two-particle transience exhibited by the original structure. In particular, we checked that the single random walk in $\mathcal{M}$ occurs to be on the line $Y_{\textrm{cm}}$ an infinite number of times, yet the probability of eventually reaching that line is strictly smaller than 1. % These results are actually derived through a proper finite-size scaling which allows to infer the behavior in the thermodynamic limit.

%Again from simulations we note that the rotation of the line of encounters
%leaves these properties unchanged, unless it coincides with the intersection
%between the binding and the page.
%The latter corresponds to on of the particles fixed VERO?

\section*{Acknowledgments}
\noindent
The authors are thankful to Alexander Blumen for interesting discussions. EA is grateful to GNFM and Sapienza Universit\`a di Roma for financial support. 

\bibliography{Omni}

\end{document}